\documentclass[fleqn,usenatbib]{mnras}

\usepackage[T1]{fontenc}
\usepackage{ae,aecompl}

\usepackage{graphicx}	% Including figure files
\usepackage{amsmath}	% Advanced maths commands
\usepackage{amssymb}	% Extra maths symbols
\usepackage{xcolor}
\defcitealias{Jauzac2015}{J15}
\title[Setting the scene for BUFFALO]{Setting the scene for BUFFALO: A study of the matter distribution in the HFF galaxy cluster MACS\,J0416.1-2403 and its parallel field}
\author[E. J. Gonzalez et al.]{
E. J. Gonzalez,$^{1,2,3}$\thanks{E-mail: ejgonzalez@unc.edu.ar (EJG)}
M. Chalela,$^{1,2}$
M. Jauzac,$^{4,5,6}$
D. Eckert,$^{7}$
M. Schaller,$^{8}$
\newauthor
D. Harvey,$^{9}$
A. Niemiec,$^{10}$
A. M. Koekemoer,$^{11}$
D. Barnes,$^{12}$
D. Clowe,$^{13}$
T. Connor,$^{14}$
\newauthor
J. M. Diego,$^{15}$
J. D. Remolina Gonzalez$^{10}$,
and C. L. Steinhardt,$^{16,17,18}$
\\
\\
$^{1}$Instituto de Astronom\'{i}a Te\'{o}rica y Experimental, (IATE-CONICET), Laprida 854, X5000BGR, C\'{o}rdoba, Argentina\\
$^{2}$Observatorio Astron\'{o}mico de C\'{o}rdoba, Universidad Nacional de C\'{o}rdoba, Laprida 854, X5000BGR, C\'{o}rdoba, Argentina\\
$^{3}$Observatorio Astron\'{o}mico de C\'{o}rdoba, Universidad Nacional de C\'{o}rdoba, Laprida 854, X5000BGR, C\'{o}rdoba, Argentina\\
$^{4}$Centre for Extragalactic Astronomy, Durham University, South Road, Durham DH1 3LE, U.K.\\
$^{5}$Institute for Computational Cosmology, Durham University, South Road, Durham DH1 3LE, U.K.\\
$^{6}$Astrophysics and Cosmology Research Unit, School of Mathematical Sciences, University of KwaZulu-Natal, Durban 4041, South Africa\\
$^{7}$Astronomy Department, University of Geneva, 16 ch. d'Ecogia, CH-1290 Versoix, Switzerland\\
$^{8}$Leiden Observatory, Leiden University, PO Box 9513, 2300 RA Leiden, The Netherland\\
$^{9}$Instituut-Lorentz for Theoretical Physics, Universiteit Leiden, Niels Bohrweg 2, Leiden, The Netherlands \\
$^{10}$Department of Astronomy, University of Michigan, 1085 South University Ave, Ann Arbor, MI 48109, USA\\
$^{11}$Space Telescope Science Institute, 3700 San Martin Dr., Baltimore, MD 21218, USA\\
$^{12}$ Department of Physics, MIT Kavli Institute for Astrophysics and Space Research, Cambridge, MA 02139, USA\\
$^{13}$Ohio University, Department of Physics and Astronomy, Clippinger Labs 251B, Athens, OH 45701, USA\\
$^{14}$The Observatories of the Carnegie Institution for Science, 813 Santa Barbara St, Pasadena, CA 91101\\
$^{15}$Instituto de F\'isica de Cantabria (CSIC-UC), Edificio Juan Jord\'a. Avda Los Castros s/n. 39005 Santander, Spain\\
$^{16}$ Cosmic Dawn Center (DAWN)\\
$^{17}$ Niels Bohr Institute, University of Copenhagen, Lyngbyvej 2, 2100 Copenhagen , Denmark\\
$^{18}$ DARK, Niels Bohr Institute, University of Copenhagen, Lyngbyvej 2, DK-2100 Copenhagen, Denmark\\
}
\date{Accepted XXX. Received YYY; in original form ZZZ}

\pubyear{2019}

\begin{document}

\label{firstpage}
\pagerange{\pageref{firstpage}--\pageref{lastpage}}
\maketitle

\begin{abstract}
In the context of the BUFFALO (Beyond Ultra-deep Frontier Fields And Legacy Observations) survey, we present a new analysis of the merging galaxy cluster MACS\,J0416.1-2403 ($z = 0.397$) and its parallel field using the data collected by the Hubble Frontier Fields (HFF) campaign. In this work, we measure the surface mass density from a weak-lensing analysis, and characterise the overall matter distribution in both the cluster and parallel fields. The surface mass distribution derived for the parallel field shows clumpy overdensities connected by filament-like structures elongated in the direction of the cluster core. We also characterise the X-ray emission of the cluster, and compare it with the lensing mass distribution. We identify five substructures at the $>5\sigma$ level over the two fields, four of them being in the cluster one. Furthermore, three of them are located close to the edges of the field of view, and border issues can significantly hamper the determination of their physical parameters. Finally, we compare our results with the predicted subhalo distribution of one of the Hydrangea/C-EAGLE simulated cluster. Significant differences are obtained suggesting the simulated cluster is at a more advanced evolutionary state than MACS\,J0416.1-2403. Our results anticipate the upcoming BUFFALO observations that will link the two HFF fields, extending further the \emph{HST} coverage, and thus allowing a better characterisation of the reported substructures.
\end{abstract}

\begin{keywords}
gravitational lensing: weak - galaxies: clusters: individual: MACS J0416.1-2403 - X-rays: galaxies: clusters - dark matter - cosmology: observations
\end{keywords}

%%%%%%%%%%%%%%%%%%%%%%%%%%%%%%%%%%%%%%%%%%%%%%%%%%

%%%%%%%%%%%%%%%%% BODY OF PAPER %%%%%%%%%%%%%%%%%%

\section{Introduction}

Massive clusters of galaxies act as natural telescopes by deflecting and magnifying the light emitted by galaxies behind them due to the gravitational lensing \citep[e.g. see reviews][]{Bartelmann2017,Kneib2010,Kneib2011,Wambsganss2006}. Taking advantage of this, the \emph{Hubble Space Telescope} (\emph{HST}) observed six of the most massive known clusters of galaxies  in the context of the Hubble Frontier Fields \citep[HFF,][]{Lotz2017} programme. The HFF combine the capabilities of \emph{HST} with the magnification power of massive galaxy clusters. The programme observed six massive strong-lensing clusters and six parallel `blank' fields (4\,arcmin away from the central field), in order to detect the faintest galaxies and to obtain hints regarding galaxy evolution at early times.

The main scientific goals of the HFF, are to explore the high redshift Universe characterising galaxies at $ z > 5 $, and to set the scene for the coming \emph{James Webb Space Telescope} (\emph{JWST}). Following a similar philosophy, the Beyond Ultra-deep Frontier Fields And Legacy Observations \citep[BUFFALO,][]{buffalo} survey expands the spatial coverage of the HFF clusters with \emph{HST} out to $3/4\times$R$_{vir}$, and covers the unobserved regions between the HFF cluster and the parallel fields. BUFFALO will place constraints on the formation of massive and luminous high-redshift galaxies as well as study how dark matter, gas and dynamics influence clusters and their surroundings. In particular, the analysis of substructures in massive clusters can be used as a test for the standard model of cosmology, $\Lambda$CDM. Detected substructures in cluster surroundings can be compared with the subhalo mass function predicted by simulations \citep[eg.,][]{Springel2001,Natarajan2004,Natarajan2007,Grillo2015,Steinhardt2016,Schwinn2017,Jauzac2018}. Moreover, comparisons between the observed and predicted radial distribution of substructures for the subhalos in simulations provides an additional test to the current cosmological paradigm.

It is important to have detailed measurements of the mass distributions of the HFF clusters in order to use them as natural telescopes. In this sense, gravitational lensing has prove to be a powerful tool to constrain the line-of-sight projected surface mass distribution of galaxy systems. Strong-lensing in particular, in which the images of source galaxies are strongly distorted and observed as arcs and multiple images, provides information on the inner regions of galaxy systems \citep[e.g. ][]{Diego2007,Zitrin2009,Vegetti2010,Lam2014,Sharon2015,Jauzac2015b,Reed2018,Williams2018,Acebron2019,Sharon2019,Mahler2019}. At the same time, weak gravitational lensing is a powerful statistical tool that provides information regarding the projected mass distribution of galaxy systems at larger distances from their centres and allows to obtain the total masses of the dark matter halos  \citep[e.g. ][]{Wegner2011,Dietrich2012,Jauzac2012,Umetsu2014,Jullo2014,Gonzalez2018}. The combination of both techniques allows us to obtain a well constrained mass distribution at small and larger distances from the cluster centres, which subsequently helps us to recover the distribution of lower-mass dark matter substructures \citep{Diego2007,Sereno2011,Jauzac2015,Jauzac2016,Jauzac2018}.

In view of the forthcoming BUFFALO observations, we present an analysis of the massive HFF cluster, MACS\,J0416.1-2403 ($z = 0.397$, hereafter MACS\,J0416). This cluster was discovered by the Massive Cluster Survey \citep[MACS; ][]{Ebeling2001}, and was classified as a merging system according to its X-ray emission \citep{Mann2012} that shows a double-peaked profile and a very elongated gas distribution. This scenario is confirmed by the strong lensing analysis presented by \citet{Jauzac2014}. Based on a set of 57 multiply-imaged systems, the best-fit model includes two cluster-scale dark matter halos, with a velocity dispersion of 778 and 955\,km\,s$^{-1}$ respectively, and 98 galaxy-scale halos. This study was extended by including weak-lensing to model the surroundings of the cluster core \citep[][ hereafter J15]{Jauzac2015} from which a third massive structure was detected in the South-West direction from the cluster centre. Despite the complex structure of MACS\,J0416 and its merging characteristics, a good correlation between mass and light is observed in this system \citep{Sebesta2016}. This cluster was also used to identify halo substructure from lensing analysis using masses lower than $ 10^{13}\,M_{\odot}$. Derived results were compared with the subhalo mass functions predicted by numerical simulations. \citet{Grillo2015} found that simulated galaxy clusters with a mass comparable to MACS\,J0416, contain considerably less mass in subhalos in their cores than the one inferred from a strong-lensing analysis. A posterior analysis found a good correlation between the predictions from simulations and the lensing infered substructures, but reported discrepancies regarding the radial distribution of the detected subhalos \citep{Natarajan2017}.

In this work we present a new optical analysis of the MACS\,J0416 cluster and parallel fields, complemented by an X-ray study of the parallel field that will be completed by the BUFFALO survey. The MACS\,J0416 parallel field was selected to lie west of the cluster in order to avoid the bright eastern stars. This orientation is perpendicular to the elongation of the cluster on the sky, so no significant mass distribution associated with the cluster is expected in this field. In this work, we pursue a new weak-lensing study of the mentioned fields, which we combine with previous strong-lensing results in order to derive the projected surface mass density of the cluster and its parallel. This approach allows to map the density distribution in the outskirts of the cluster and to detect the presence of subhalos, since the strong lensing information sets the location and shape of the cluster core while the weak lensing mimics the mass distribution at larger scales. The resulting surface distribution is then put in perspective of the optical and X-ray emission distributions. From the derived lensing projected mass distribution, we also identify substructures present in these fields and compare our results with predictions from numerical simulations. 

The paper is organised as follows. In Sec.\,\ref{sec:data} and Sec. \ref{sec:simu}, we describe the observational and simulated data used to perform the analysis respectively. In Sec.\,\ref{sec:catalogues}, we describe the shape measurements and define the criterion for the galaxy classification for the weak-lensing analysis. In Sec.\,\ref{sec:model}, we characterise the method used to obtain the projected mass distribution from the shape measurements. We present our results in Section\,\ref{sec:results}. Finally, we discuss our results in Sec. \ref{sec:conclusions}. Throughout the analysis, we adopt a standard cosmological model: $H_0 = 70$\,km\,s$^{-1}$\,Mpc, $\Omega_m = 0.3,$, $\Omega_\Lambda = 0.7$. All magnitudes are quoted in the AB system.

\section{Observations}
\label{sec:data}
\subsection{\emph{Hubble Space Telescope}}
MACS\,J0416 was first observed using \textit{HST} in 2007 under the SNAPshot programme GO-11103 (PI: Ebeling) using the Wide Field and Planetary Camera 2 (WFPC2). These observations pointed at MACS\,J0416 as a powerful lens, which led to the inclusion of this cluster in the CLASH programme \citep[PI: Postman;][]{Postman2012}. The cluster was observed again in 2012 for a total of 20 orbits across 16 passbands, from the UV to the near-IR. The obtained data were used for the pre-HFF analysis of the cluster \footnote{All published mass models based on the pre-HFF data by \citet{Coe2015,Johnson2014,Richard2014} are publicly available at \href{http://archive.stsci.edu/prepds/frontier/lensmodels/}{http://archive.stsci.edu/prepds/frontier/lensmodels/}}. More information regarding these images can be found in \citetalias{Jauzac2015}.

For the lensing analysis we use the same dataset as the one described in \citetalias{Jauzac2015}, based on the observations taken with the Advanced Camera for Surveys (ACS). We also include in our analysis the parallel `blank' field images in the same filters as for the cluster field ($F435W$ $F606W$ and $F814W$). All of these HFF observations were performed under the observing programme GO-13496 \citep[PI: Lotz,][]{Lotz2017}. 

Reduced images were obtained after applying basic data-reduction procedures, using \textsc{hstcal} and the most recent calibration files. Individual frames were co-added for each filter, using \textsc{astrodrizzle} after registration to a common ACS reference image using \textsc{tweakreg}. \textsc{astrodrizzle} generates the drizzled images, correcting for the geometric distortion that is produced since ACS is located off-axis in the \emph{HST} focal plane and the
ACS focal plane is not normal to incident light rays. This is done simultaneously removing cosmic rays and bad pixels, as well as combining multiple exposures into a single output image. Final stacked images have a pixel size of 0.03\arcsec. A summary of the observations and some of their characteristics are provided in Table \ref{tab:observations}.

\begin{table*}
\caption{Summary of the \emph{HST} observations used in this work. 
}
    \label{tab:observations}
    \centering
    \begin{tabular}{c c c c  c c c}
        \hline
        \hline
         Field & RA (J2000) & Dec. (J2000) & Number of & Date range & Instrument/Filter & Exposure time \\ 
         & & & combined exposures &  of observations & & (in seconds)\\
        \hline
        Cluster  & 04:16:08.9 & $-$24:04:28.7 & 40 & 2014-02-21/22 & ACS/$F435W$ & 54\,512 \\
                 & &  & 24 & & ACS/$F606W$ & 33\,494 \\
                 &  &  & 96 & & ACS/$F814W$ & 129\,941 \\
        Parallel & 04:16:33.1 & $-$24:06:50.6 & 36 & 2014-09-05 & ACS/$F435W$ & 45\,747 \\
                 &  &  & 20 & & ACS/$F606W$ & 25\,035 \\
                 &  &  & 83 & & ACS/$F814W$ & 105\,498 \\
        \hline
\end{tabular}
\medskip
\begin{flushleft}
\end{flushleft}    
\end{table*}

\subsection{Spectroscopic and photometric redshifts}

We make use of the HFF-DeepSpace photometric catalogues of the twelve HFF presented in \citet{Shipley2018}. These catalogues were constructed using all data publicly available from space and ground-based observations. These include \emph{HST}/WFC3, \emph{HST}/ACS, \emph{Spitzer Space Observatory}/IRAC, the \emph{Very Large Telescope}/HAWK-I, and \emph{Keck-I}/MOSFIRE, providing a total of 22 filters for photometry, and thus photometric redshifts of excellent quality. Photometric redshifts were computed with the \texttt{EAZY} software \citep{Brammer2008}. To asses their quality, all spectroscopic redshifts available in the literature were used (only from sources that targeted the \textit{HFF} clusters), achieving an average scatter of $\sigma \sim 0.034$ between photometric and spectroscopic redshifts. These spectroscopic redshifts are also included in the HFF-DeepSpace catalogues. 

For the cluster and parallel fields of MACS\,J0416 there are 378 and 79 spectroscopic redshifts respectively, listed from different sources:  \citet{Jauzac2014,Ebeling2014,Grillo2015}, GLASS \citep{Treu2015};\citet{Balestra2016,Caminha2017} and Brammer et al. (in prep.). All of these redshifts, as well as the photometry provided in the HFF-DeepSpace catalogues, in particular the $F435W$, $F606W$ and $F814W$ pass-bands corrected for galactic extinction, are used in our analysis for the selection of background galaxies, and the identification of cluster members. Following the prescriptions of \citet{Shipley2018}, we restrict the galaxies used in this work to those with flag use$\_$phot=1 and with a strict cut in the photometry signal-to-noise ratio of S/N>10 \citep[further details on these parameters can be found in Sec. 3.10 of ][]{Shipley2018}. 

\subsection{\emph{Chandra X-ray Observatory}}

We compare the derived surface mass density distribution derived for the parallel field with the X-ray emission using the X-ray data provided by Chandra. MACS\,J0416 was observed by \emph{Chandra}/ACIS-I on six occasions between 2009 June and 2014 December (observation ID 10446, 16236, 16237, 16304, 16523, and 17313) for a total of 324 ks. The full dataset was analyzed in detail in \citet{Ogrean2015}. We reprocessed the six individual observations using the \texttt{CIAO} v4.8 package and CALDB v.4.7.2 with the \texttt{chandra\_repro} tool. We inspected the light curves of each individual observation to remove periods of flaring background and create clean event files. The individual event files were then merged using the \texttt{merge\_obs} utility. We extracted images and exposure maps in the [0.5-2] keV energy band from the merged event files using \texttt{fluximage} tool. Finally, we used a collection of blank-sky images to estimate a local background map by reprojecting the events along the telescope's attitude \citep{Hickox2006}. 

\section{Simulations}
\label{sec:simu}

From the lensing surface mass density distribution, we detect subhalos present in the cluster and parallel fields. The derived substructures and their distribution are compared with simulations. 
We perform a similar analysis as the one presented in \citet{Jauzac2018} by comparing our lensing detected subhalos with the ones detected in a simulated cluster similar to MACS\,J0416. In order to do that, we use the Hydrangea/C-EAGLE simulation \citep{Bahe2017,Barnes2017}, a set of cosmological hydrodynamical zoom-in simulations of the formation of 30 galaxy clusters in the mass range $10^{14} < M_{200}/M_\odot < 10^{15}$. 
The clusters were selected from a parent, dark matter only simulation of 3.2 Gpc length-side \citep{Barnes2017} based on the cosmological parameters derived from the 2013 analysis of the Planck data \citep[$H_{0}=100{\rm h}=67.77$\,km\,s$^{−1}$\,Mpc$^{−1}$, $\Omega_{\Lambda}=0.693$, $\Omega_{m} = 0.307$, $\Omega_{b}= 0.04825$, $\sigma_{8} = 0.8288$, $n_s=  0.9611$, and $Y =  0.248$;][]{Planck2014}.

Dark matter halos were selected using the Friends-of-Friends algorithm \citep{Davis1985}, and bound subhalos were identified using the \textsc{subfind} algorithm \citep{Springel2001,Dolag2009}. Thirty halos at $z = 0$ were selected for the zoom-im realisation taking into account a mass $M_{200} > 10^{14} M_\odot$ and an isolation criterion (no other massive halos within 20 times the $R_{200}$ radius). 

The EAGLE simulation code \citep{Schaye2015,Crain2015,Schaller2015} is used to resimulate the halo selection sample assuming a mass of $9.7 \times 10^{6}\,M_\odot$ and $1.8 \times 10^{6}\,M_\odot$ for the dark matter and gas particles respectively, a softening length of $2.66$ comoving kpc for $z>2.8$ and a physical softening length of $0.70$\,kpc for $z<2.8$. Post-processed halo and sub-halo catalogues were generated for all output redshifts using the \textsc{subfind} algorithm. Simulated clusters attempt to reproduce the formation of rich galaxy clusters with a model that yields a galaxy population that is a good match to the observed field population. 

\section{Galaxy catalogues}
\label{sec:catalogues}

In this section we detail the galaxy catalogues used for the lensing analysis and the optical luminosity distribution. We first present the source detection, photometry and shape measurements performed in the optical stacked images described in Sec. \ref{sec:data}. Then we discuss the background source identification. Cluster members are used to model the cluster gravitational potential and to obtain the optical light distribution. Background galaxies, defined as galaxies behind MACS\,J0416 and hence lensed by the cluster, are used to perform the weak-lensing analysis. 

\subsection{Source catalogue and shape measurements}
In order to detect sources in the HFF images, and measure the shapes of background galaxies for the weak-lensing analysis, we use the ACS/$F814W$ filter. As in \citetalias{Jauzac2015}, we follow the same approach as for the COSMOS survey \citep{Leauthaud2007} when adapted to cluster fields \citep{Jauzac2012}. We compute the shapes using the pipeline \texttt{pyRRG}\footnote{https://github.com/davidharvey1986/pyRRG} developed by D. Harvey \citep{harvey15,harvey19} and based on the RRG method \citep{Rhodes2000}.

The source detection and the photometry are performed using the \texttt{SExtractor} package \citep{Bertin1996} applying the `\textit{Hot-Cold}' method: (1) \texttt{SExtractor} is executed with a configuration optimised to detect only the brightest objects (\textit{cold} step), and (2) \texttt{SExtractor} is run with a configuration optimised to detect faint objects (\textit{hot} step). The two catalogues are then merged by including all objects detected during the \textit{cold} step, plus the objects detected during the \textit{hot} step but not the \textit{cold} step. Finally, double detections are removed by discarding all objects within one FWHM\_IMAGE of each other, keeping larger objects.
The source classification as galaxies, stars and false detections is performed according to the distribution of the objects in the MAG\_AUTO
\emph{versus} peak surface brightness MU\_MAX plane \citep[see][for further details]{Leauthaud2007}.

Galaxy shapes are computed using the RRG method \citep{Rhodes2000}. This method was specifically developed for weak-lensing analysis of space-based observations. Since the ACS Point Spread Function (PSF) varies due to the telescope `breathing', the effective focus of the observation is determined by comparing the ellipticity of the sources classified as stars with a grid of simulated PSF images generated by \citet{Rhodes2007}. PSF parameters are interpolated first creating a grid of positions which covers the entire field-of-view (FOV) of the combined drizzled image. Then, for each position in the drizzled image, the \texttt{pyRRG} code identifies how many images cover this position and computes the PSF while rotating the moments such that they are in the reference frame of the stacked image. Finally, it averages the moments over the stack to obtain the PSF at the considered position.

From the RRG method we obtain the galaxy moments corrected from instrumental effects. For each galaxy, the ellipticity, $e = (e_1,e_2)$, and the size, $d$, are computed as:
\begin{equation*}
    e_1 = \frac{I_{xx} - I_{yy}}{I_{xx} + I_{yy}}\ ,
\end{equation*}
\begin{equation}
    e_2 = \frac{2I_{xy}}{I_{xx} + I_{yy}}\ ,
\end{equation}
\begin{equation*}
    d = \sqrt{\frac{I_{xx} + I_{yy}}{2}}\ ,
\end{equation*}
where $I_{ij}$ are the second order Gaussian-weighted moments. The shear estimator, $\tilde{{\gamma}}$, is obtained from the measured ellipticities according to:
\begin{equation}
    \tilde{{\gamma}} = C \frac{e}{G}\ ,
\end{equation}
where $G$ is the shear susceptibility and is computed following equation\,28 in \citet{Rhodes2000}, and $C$ is the calibration factor \citep[$C = 0.86$, see][for further details]{Leauthaud2007}.

Finally we only consider galaxies with $22.5 < m_{F814W} < 30.0$ for our weak-lensing analysis. We also discard galaxies with shape measurements based on fewer than three exposures, in order to discard galaxies near the edges of the observed fields. We also set a threshold on the detection limit ($S/N =$ FLUX\_AUTO/FLUXERR\_AUTO $ > 4.5$),  the ellipticity parameter ($e < 1$), and the size ($3.6 < d < 30$ pixels).

\subsection{Cluster members}
\label{subsec:members}
The contribution of cluster galaxies must be carefully considered in the mass modelling based on strong lensing observations \citep{clusterMembers}. In this work we perform a weak lensing analysis combined with the best-fit strong-lensing mass model from previous work (see Sec. \ref{sec:model}) to model the projected mass density distribution in the outskirts of the cluster, where cluster member contributions to the total mass distribution is not significant. Therefore, the inclusion of interlopers in the cluster member sample has not a major impact on the mass modelling. Moreover, since these galaxies are considered for the background galaxy selection, we adopt a relax criterion for their identification. 

\begin{figure}
    \centering
    \includegraphics[scale = 0.45]{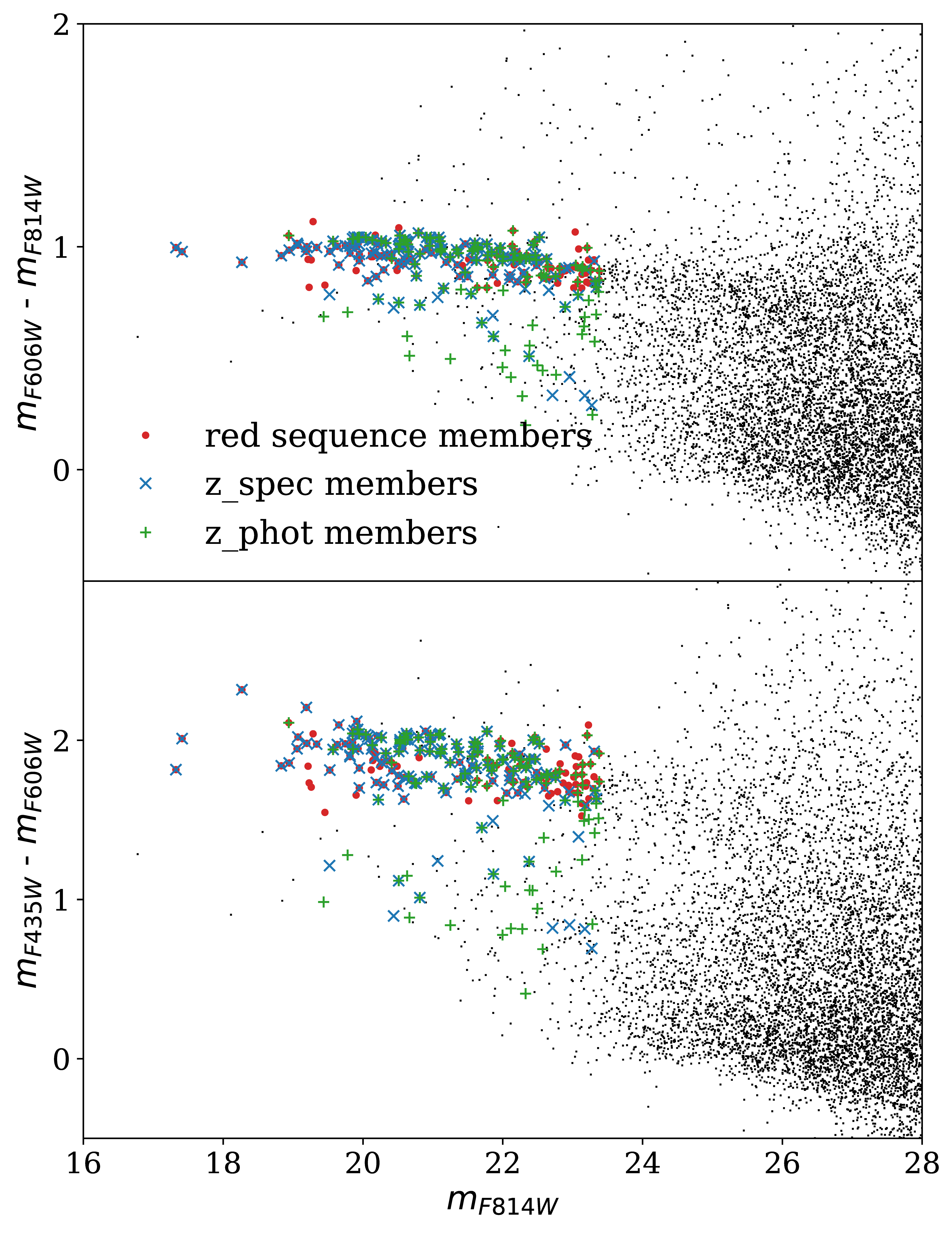}
    \caption{Colour-magnitude diagrams for all galaxies within the cluster and the parallel fields. Red sequence galaxies are selected according to a gaussian fit, including all the galaxies within 3$\sigma$, and are marked as red dots. Galaxies selected according to their photometric and spectroscopic redshifts are marked as blue crosses and green pluses, respectively. }
    \label{fig:members}
\end{figure}

We identify cluster members in each field following the criteria presented in \citetalias{Jauzac2015}. For both cluster and parallel fields, we identify cluster members as galaxies with photometric redshifts ($z_{phot}$ members) that satisfy $0.35 < z_{phot} < 0.44$, and those with spectroscopic redshifts ($z_{sepc}$ members) that satisfy $ \mid z_{cluster} - z_{spec} \mid < 0.0104$, 
 with $z_{cluster} = 0.3979$. We also include galaxies that fall within 3$\sigma$ of the cluster red-sequence in both ($m_{F606W} - m_{F814W}$) versus $m_{F814W}$ and ($m_{F435W} - m_{F606W}$) versus $m_{F814W}$ colour-magnitude diagrams (red-sequence members). We model the red-sequence with a Gaussian function, fitting the colour-magnitude distribution of galaxies with $m_{F814W} > 22$. In spite this being a rough selection which can lead to the inclusion of non-cluster members \citep{Connor2019}, all these galaxies are taken into account in order to place more conservative boundaries for the background galaxy selection. The mean of the best-fit Gaussian is 0.99 (1.92) with a standard deviation of 0.06 (0.14) for the galaxies in the cluster (parallel) field.  In total we identify 245 galaxies as cluster members. In Fig.~\ref{fig:members} we show both colour-magnitude diagrams with the selected members marked. 
 
\subsection{Background galaxies}
\label{subsec:sources}
We select background galaxies, defined as galaxies behind the cluster and lensed by it, following \citetalias{Jauzac2015}. We take into account the position of the sources classified as galaxies in the colour-colour space. 

For galaxies with either spectroscopic or photometric redshifts, we classify them as foreground galaxies if $z_{spec} < z_{cluster} - dz$ and $z_{phot} < 0.35$. According to this classification and considering cluster members identified previously, we identify a region in the colour-colour space defined as: 
$$(m_{F435W} - m_{F606W}) > 0.3 \ ,$$
$$(m_{F435W} - m_{F606W}) < 0.67776\,,\,(m_{F435W} - m_{F814W}) + 0.2 \ ,$$
$$(m_{F435W} - m_{F606W}) > 0.87776\,,\,(m_{F435W} - m_{F814W})-0.86\ .$$

\begin{figure}
    \centering
    \includegraphics[scale = 0.45]{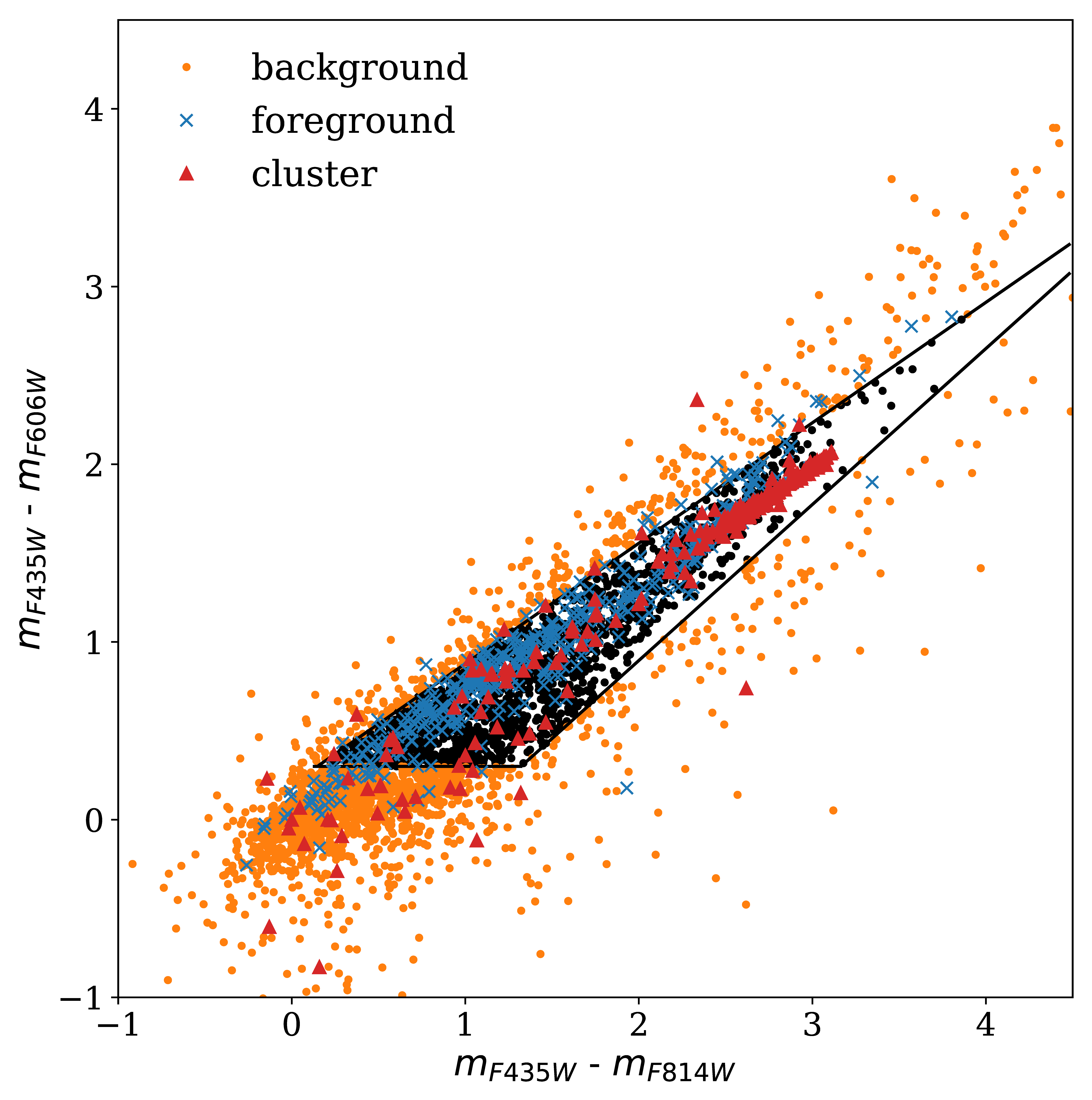}
    \caption{Colour-colour diagram for all galaxies detected within the cluster and the parallel fields. Galaxies classified as foreground and cluster members are marked with crosses and triangles, respectively. Black lines mark the background galaxy selection region. Galaxies outside that region are considered as background objects (orange points).}
    \label{fig:background}
\end{figure}

\begin{figure}
    \centering
    \includegraphics[scale = 0.5]{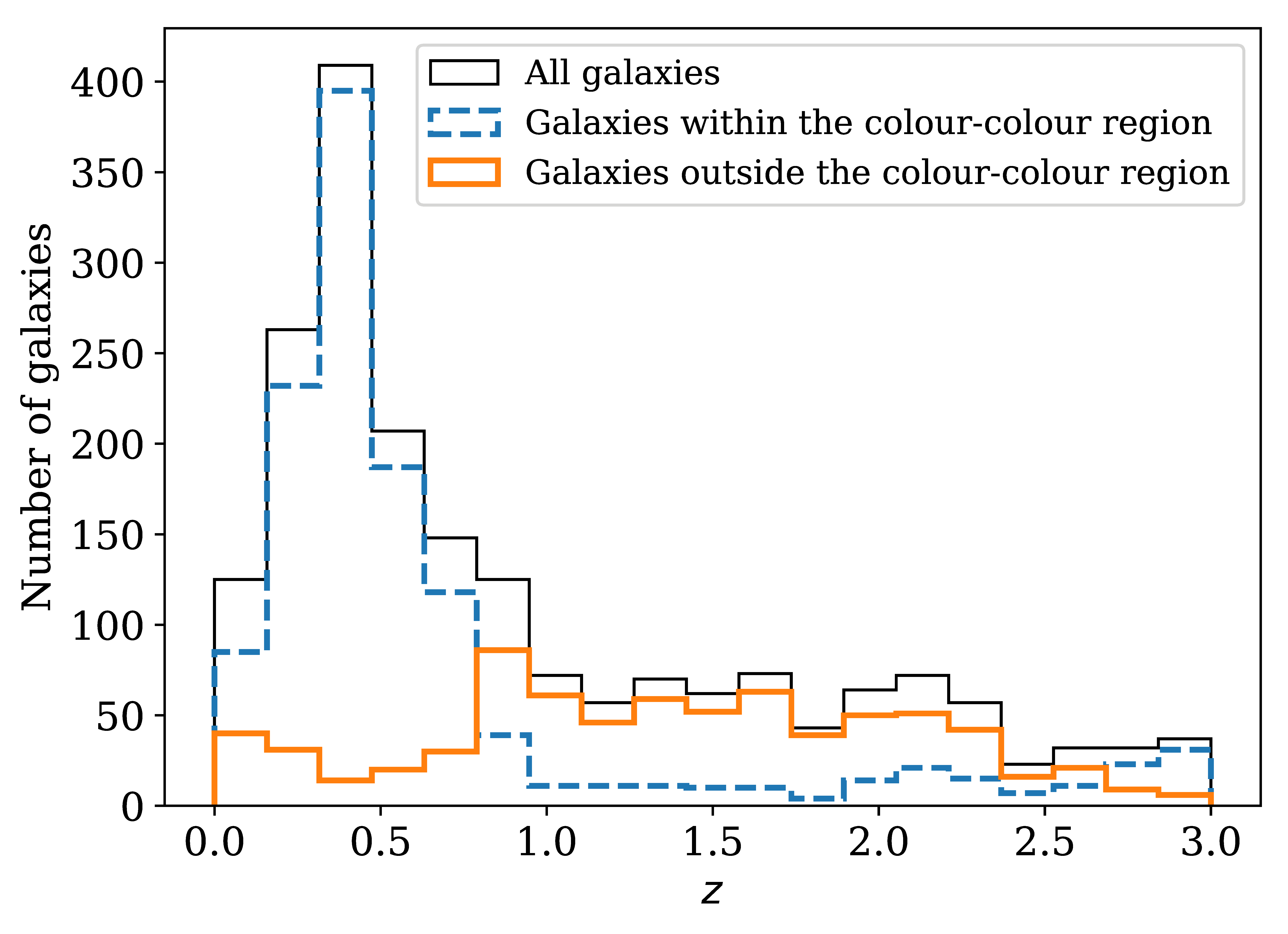}
    \caption{Redshift distribution of all galaxies that have photometric or spectroscopic redshifts (solid line). The dashed histogram and the thicker
    line histogram show the distribution of galaxies within and outside the colour-colour region defined for the background galaxy selection.}
    \label{fig:zdist}
\end{figure}

All galaxies within this region are considered as either foreground or cluster objects. They are thus removed from our final weak-lensing catalogue together with galaxies at $z_{phot} > 3$ (see Fig. \ref{fig:background}). In Fig. \ref{fig:zdist} we show the redshift distribution for the subset of galaxies with redshift information that lie within and outside the defined colour-colour region. Approximately $\sim 89 \%$ of the unlensed galaxies (foreground and cluster) are discarded using this colour-colour criterion. 

We classify 1684 sources as background galaxies, 549 of which have redshift information. With this selection criteria we obtain a background galaxy density of $\sim 70$ and $\sim 50$ galaxies\,arcmin$^{-2}$ for the cluster and parallel field, respectively. These differences in the observed galaxy density are mainly due to the shorter exposure for the frames observed in the parallel field (see Table \ref{tab:observations}) as well as for the larger density of member galaxies in the cluster field that hamper the detection of fainter sources.

To model the redshift distribution for the galaxies without redshift information, we use the following function:
\begin{equation}
    N(z) = z^{\alpha} \exp{-(z/z_0)^{\beta}}.
\end{equation}
We fix $\alpha = 2$, and fit the redshift distribution of the background galaxies with redshift information, obtaining $z_0 = 1.28$ and $\beta = 1.82$. 

\section{Mass Modelling}
\label{sec:model}

We model the mass distribution using a grid-based model combined with a parametric model following a similar approach as in \citetalias{Jauzac2015}. The grid-based model consists of a set of radial basis functions (RBFs) located at the nodes of the multiscale grid \citep{Jullo2009,Jullo2014}. Each RBF is modelled with a dual pseudo isothermal elliptical mass distribution \citep[dPIE,][]{Eliasdottir2007}. This profile is a two component pseudo-isothermal mass distribution with a core radius ($r_c$, defined as the distance between an RBF and its closest neighbour) and a scale radius \citep[assumed to be three times $r_c$,][]{Jullo2009}. During the fitting procedure only the potential amplitudes vary according to the lensing signal and hence, follows the projected density distribution. 
We use a uniform grid of the same size as the FOV. The resolution of the grid is given by the core radius parameter, $r_c$, which also could be understood as a softness parameter. 

We select this parameter considering the computed signal-to-noise ratio (S/N) maps. We tried three different grid resolutions, $3.0\arcsec$, $4.5\arcsec$ and $6.0\arcsec$ for the cluster field and $6.0\arcsec$, $7.5\arcsec$ and $9.0\arcsec$ for the parallel field, since a lower signal is expected for this field. No significant differences were obtained in the resultant mass density distributions derived for the different resolutions considered. Nevertheless, we fix $r_c$ at $6\arcsec$ and $9\arcsec$ for the cluster and parallel fields, respectively, since for lower resolutions the obtained distributions show clumpy structures with S/N lower than 3. On the other hand, for the main field larger resolution grid, substructures detected in the surface density map at a high significance (S/N > 5) are merged.
At this resolution, the obtained grids consist of 1200 and 770 nodes for the cluster and parallel field respectively.

For the cluster field, we combine the grid with two cluster-scale dark matter halos and 219 cluster members in the inner core, optimised using strong-lensing constraints. These two dark matter halos are modelled with pseudo-isothermal elliptical mass distributions \citep[PIEMD;][]{Eliasdottir2007}, according to the results from the strong-lensing analysis presented in \citet{Jauzac2014} (Table\,\ref{tab:halos}). 

On the other hand, the parallel field does not include large-scale dark matter halos as in the core. We only include the 26 cluster members galaxies identified as described in Sect.~\ref{subsec:members}. Those are also modelled with dPIE potentials with parameters fixed according to \citetalias{Jauzac2015}: $m^{*} = 19.76$, $\sigma^{*} = 119$\,km\,s$^{-1}$ and $r^{*}_{cut} = 85$\,kpc. Although these parameter are derived for galaxy members close to the cluster core, we do not expect significant differences in the derived surface density mass.

\begin{table}
\caption{Derived pseudo-isothermal elliptical mass distribution parameters according to the strong-lensing analysis presented in \citetalias{Jauzac2015}. The corresponding parameters describe the two cluster scale halo components (C1 and C2) of the cluster.}
    \label{tab:halos}
    \centering
    \begin{tabular}{c c c  }
        \hline
        \hline
         Component & C1 & C2\\ 
         \hline
        RA (J2000)  & 04:16:09.4 & 04:16:07.5 \\
        Dec. (J2000) & $-$24:04:01.4 & $-$24:04:47.4  \\
        $e$ & 0.7 & 0.7 \\
        $\theta$ & 148.0 & 127.4 \\
        $r_{core}$ (kpc) & 77.8 & 103.3 \\
         $r_{cut}$ (kpc) & 1000 & 1000 \\
         $\sigma$ (km\,s$^{-1}$) & 779 & 955 \\
        \hline
\end{tabular}
\medskip
\begin{flushleft}
\end{flushleft}    
\end{table}

While the parametric model used to trace the core projected density distribution is fixed to the best-fit obtained by J14, the RBFs are optimized using the weak-lensing constraints identified in Sect.~\ref{sec:catalogues}  in both fields individually. According to the selection described in Sect.~\ref{subsec:sources}, we use 984 and 700 background galaxies for the cluster and parallel fields respectively. To implement this fitting procedure, we use \texttt{lenstool}\footnote{https://projets.lam.fr/projects/lenstool/wiki} \citep{Jullo2007} which includes a Bayesian optimisation based on \href{BayeSys}{http://www.inference.org.uk/bayesys/}. The projected mass for each field is obtained by averaging the results of the 200 iterations, and errors are based on the standard deviation of the derived maps.

\section{Results}
\label{sec:results}
In this section we present the derived projected surface density maps reconstructed from the lensing analysis and compare it with the optical and X-ray luminosity distribution. In order to do that, we compute the optical luminosity maps of both the cluster and the parallel fields by pixellating the FOV and adding the brightness of the enclosed cluster members in each pixel. For this we compute the brightness according to their magnitude in the $F814W$ pass-band, $m_{F814W}$. We then smooth the brightness map using a Gaussian kernel with a standard deviation of 7.56\arcsec and 27\arcsec for the cluster and the parallel fields respectively. Brightness and projected mass contours are obtained using \textsc{SAOImage DS9}\footnote{\href{http://ds9.si.edu/site/Home.html}{http://ds9.si.edu/site/Home.html}}.

According to the surface mass density maps obtained with our lensing reconstruction, we detect five substructures with significance $> 5\sigma$ in the cluster and parallel fields , i.e. five times the median S/N in an aperture of 10\arcsec centred on each detected overdensity. In Table\,\ref{tab:structures}, we describe the properties of all the substructures detected in this work within the cluster and the parallel fields, together with the substructures detected in the cluster field by \citetalias{Jauzac2015}. To discard that detected substructures are produced by outliers in the background galaxy sample with less reliable shear measurements or by the inclusion of artifacts with abnormally high ellipticity, we recompute the surface mass density maps but randomly discarding $10\%$ of the background galaxies. We perform 100 realisations for both the cluster and parallel fields. We then measure the mass in fixed apertures at the locations of the substructures. The distributions of the computed masses for each of them are normal-behaved with a dispersion comparable to the estimated errors presented in Table\,\ref{tab:structures}. Therefore, we discard that the detected overdensities can be produced by outliers in the galaxy sample. 

In the next subsections we discuss the results for the cluster and parallel field separately, comparing our mass estimations with previous analysis.

\begin{table*}
\caption{Properties of the detected substructures within the main and the parallel fields. S1 and S2 are the overdensities detected and described by \citetalias{Jauzac2015}. S1c, S2c, S3c and S4c are the substructures detected in this work in the cluster field and S1p is the substructure detected in the parallel field. Masses are in units of $ 10^{13}$\,h$_{70}^{-1}$\,M$_{\odot}$.}
    \label{tab:structures}
    \centering
    \begin{tabular}{c c c c c c c}
        \hline
        \hline
         ID & RA (J2000) & Dec. (J2000) & $M$($R<100$\,kpc) & $M$($R<200$\,kpc) & $\sigma$ & $D_{G1-S}$ [kpc]\\ 
         \hline
        S1  & 4:16:03.970 & $-$24:05:41.66 & $4.2 \pm 0.6$ & -  & 7.5 & 650 \\
        S2  & 4:16:14.633 & $-$24:03:49.09 & $1.5 \pm 0.2$ & -  & 7.3 & 409 \\
        S1c & 4:16:02.770 & $-$24:03:49.09 & $3.7 \pm 0.6$ & $11.1 \pm 1.5$ & 5.8 & 687 \\
        S2c & 4:16:03.885 & $-$24:04:33.48 & $3.6 \pm 0.4$ & $12.5 \pm 1.3$ & 8.4 & 418 \\
        S3c & 4:16:07.571 & $-$24:02:56.10 & $2.8 \pm 0.3$ & $8.6 \pm 1.2$ & 8.7 & 376 \\
        S4c & 4:16:14.746 & $-$24:03:13.96 & $3.5 \pm 0.5$ & $11.3 \pm 2.1$ & 7.0 & 487 \\
        S1p & 4:16:32.207 & $-$24:05:18.33 & $2.1 \pm 0.4$ & $5.0 \pm 1.1$ & 6.3 & 1737 \\
        \hline
\end{tabular}
\medskip
\begin{flushleft}
\end{flushleft}    
\end{table*}

\subsection{Cluster field}

In Fig. \ref{fig:main} we show the composite colour \emph{HST}/ACS images in the $F814W$, $F606W$ and $F435W$ pass-bands together with the surface mass density (solid lines) and brightness contours (dashed lines) for the cluster field. Surface density contours are obtained from ($1.17 \times 10^{9}$) up to ($2.93 \times 10^{9}$)\,h$_{70}^{-1}$\,M$_{\odot}$\,kpc$^{-2}$. Brightness contours are obtained from $m_{F814W} = 23.3$ up to  $m_{F814W} = 20.0$. There is a good agreement between the projected mass and the brightness distributions. 

We derive projected masses within circular apertures centred at the brightest galaxy member location (G1: RA\,(J2000) = 4:16:09.144, DEC\,(J2000) = $-$24:04:02.94) considering different aperture radii in order to compare our results with previous mass determinations. Taking into account an aperture of $R<200$\,kpc, we obtain $M$($R<200$\,kpc) $= (1.93 \pm 0.07)\times 10^{14}$\,h$_{70}^{-1}$\,M$_{\odot}$. This value is higher than other previous mass determinations as $(1.63 \pm 0.03)\times 10^{14}$\,h$_{70}^{-1}$\,M$_{\odot}$, $(1.66 \pm 0.05)\times 10^{14}$\,h$_{70}^{-1}$\,M$_{\odot}$ and  between 1.72 and $1.77\times 10^{14}$\,h$_{70}^{-1}$\,M$_{\odot}$ obtained by \citet{Richard2014}, \citetalias{Jauzac2015} and \citet{Grillo2015}, respectively. We obtain $M$($R<250$\,kpc) $= (2.71 \pm 0.12)\times 10^{14}$\,h$_{70}^{-1}$\,M$_{\odot}$, which is also higher than what has been reported by other authors: between  2.35 and 2.43$ \times 10^{14}$\,h$_{70}^{-1}$\,M$_{\odot}$ according to \citet{Grillo2015} and significantly higher than the reported masses by \citet{Johnson2014} and \citet{Gruen2014} of $(1.8 \pm 0.3)\times 10^{14}$\,h$_{70}^{-1}$\,M$_{\odot}$ and $(1.77^{+0.31}_{-0.13})\times 10^{14}$\,h$_{70}^{-1}$\,M$_{\odot}$, respectively. Finally we obtain $M$($R<320$\,kpc) $= (3.92 \pm 0.22)\times 10^{14}$\,h$_{70}^{-1}$\,M$_{\odot}$, higher than the reported values by \citet{Jauzac2014, Jauzac2015, Grillo2015}. As stated by \citet{Grillo2015}, differences in these mass estimates could be caused by displacements in the adopted cluster mass centres, details of the lensing models, and/or the degeneracy between the mass of a lens and the redshift of a multiply imaged source. Also, in contrast to the results obtained in \citetalias{Jauzac2015}, the inclusion of weak-lensing information in this work led to larger mass estimates. Differences could be due to the new redshift information included in this work.

\citetalias{Jauzac2015} identified two substructures close to the cluster, S1 and S2, which are marked in Fig. \ref{fig:main}, and described in Table \ref{tab:structures}. S1 is also confirmed by the presence of a galaxy overdensity in this region coincident with a peak in the light distribution. Although there is no X-ray emission excess confirmed in this region, the overall cluster emission is elongated in the direction of both mass structures of the cluster core \citep{Ogrean2015}. Moreover, if the substructures have already merged with the cluster halos, the dark matter could be decoupled from the gas and, therefore, a X-ray remnant core would not be expected. This scenario is also supported for S1 since \citet{Ogrean2015} reported a density discontinuity close to this substructure that could be originated by a previous interaction between one of the main halo components, C2, and S1. 

In this work, we detect four substructures in the cluster field, labelled as S1c, S2c, S3c and S4c. These are marked in Fig. \ref{fig:main} and their properties are detailed in Table \ref{tab:structures}. S1c and S2c are located close to galaxy overdensities which support their detections in our lensing analysis. Also, S1c is located close to S1, and has a projected mass of $M$($R<100$\,kpc) $= (3.74 \pm 0.64)\times 10^{13}$\,h$_{70}^{-1}$\,M$_{\odot}$, which is in agreement with the \citetalias{Jauzac2015} estimate for S1 ($M$($R<100$\,kpc) $= (4.22 \pm 0.56)\times 10^{13}$\,h$_{70}^{-1}$\,M$_{\odot}$). Moreover, there is an excess in the X-ray emission close to the location of S2c that can be observed in the X-ray map presented by \citet{Ogrean2015} \citep[see Fig. 8 in ][]{Ogrean2015}. On the other hand, S3c has no counterpart in the brightness map. Nevertheless, it is detected with high significance and could possibly correspond to a dark matter halo that already interacted with the cluster. To test if the detected substructures are not caused by the modelling considerations, we compute the projected surface density maps using only the grid and neglecting the parametric contribution of the two main halos and the galaxy members. Considering this analysis we do not detect significant signal close to S3c and S4c locations, therefore these substructures can be produced by the imposition of the parametric model on the lensing data. We also perform a quick test by obtaining the surface distribution using the reconstruction method developed by \citet{Kaiser1993}. With this analysis, we obtain a significant density distribution at the two main halo locations, C1 and C2, and close to Sc1 and Sc2. Further inspection about how the modelling can impact the detection of substructures, which is out of the scope of this paper, needs to be performed in order to asses for these discrepancies. 

\begin{figure*}
    \centering
    \includegraphics[scale = 0.6]{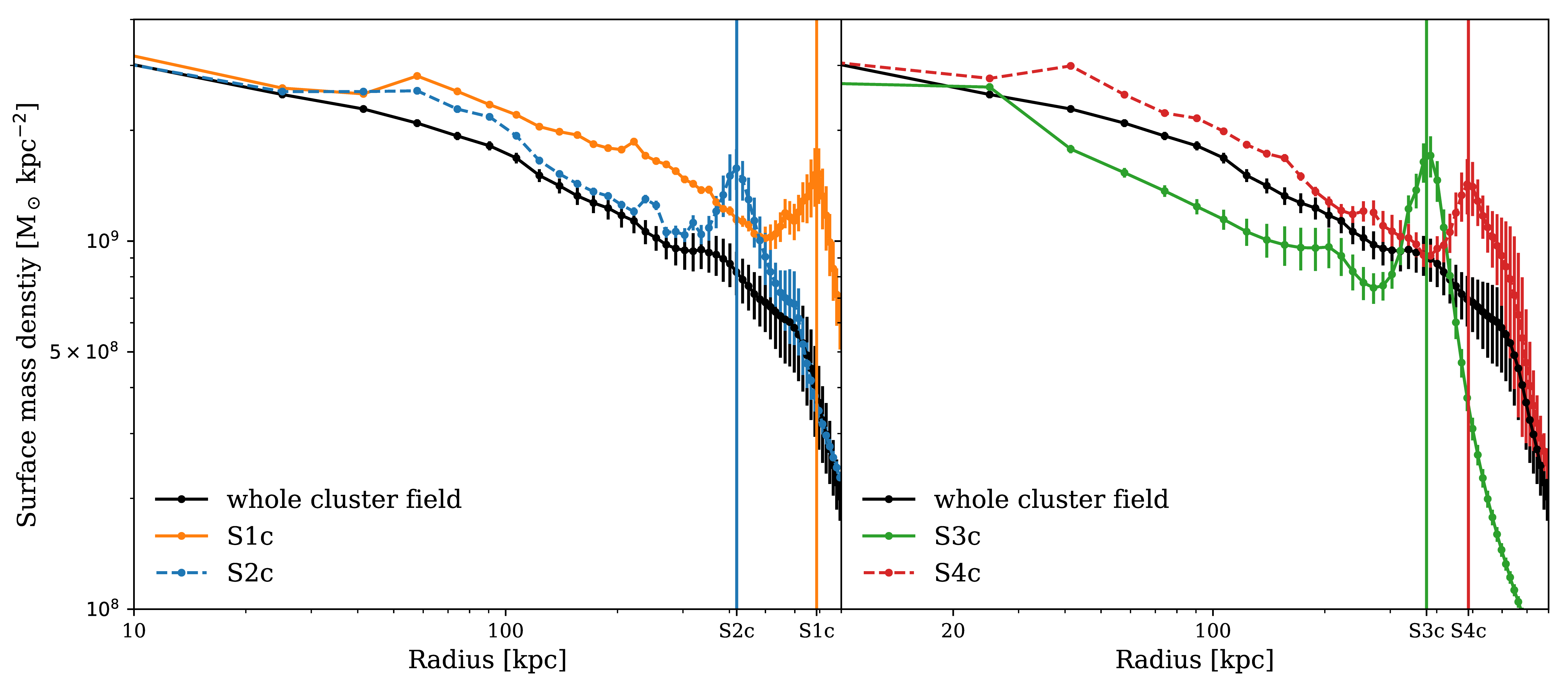}
    \caption{Surface mass density profiles obtained for the complete cluster field (solid line). Dashed lighter lines show the profiles computed within a triangular wedge region, defined with an amplitude of 10$\deg$ with their apex located at G1 and pointing to S1c and S2c substructure (left panel) and to S3c and S4c (right panel). Substructures are detected as an excess in the surface mass density at the locations of each overdensity.}
    \label{fig:profile}
\end{figure*}

There is no significant signal close to the location of S2. Nevertheless, if we lower the threshold in the mass contours, we can detect an overdensity close to this substructure but with a detection significance threshold lower than $5\sigma$. This is comparable with the detection level of some artefacts detected close to the edge of the field. Edge effects significantly hamper the identification of substructures in these regions. Errors in surface density maps start to significantly increase at distances larger than $\sim 350\,kpc$, with the median error at the edges ($350-550\,kpc$) more than two times higher than in the central region of the cluser field. Such discrepancies could be addressed by the BUFFALO survey (GO-15117; PIs: Steinhardt \& Jauzac) which images the surrounding area of the HFF cluster field of MACS\,J0416.

In Fig. \ref{fig:profile}, we show the surface density profile computed using the projected lensing mass map for the cluster field, and centered on the brightest galaxy member, G1. We also compute profiles centred in G1 but restricting the field to a triangular wedge aperture of 10\,deg with their apex located at G1 and pointing in the direction of each detected substructure. As one can see, the presence of these overdensities can be detected in the profiles as peaks located at their respective distances from G1, $D_{G1-S}$, detailed in Table \ref{tab:structures}. 

\begin{figure*}
    \centering
    \includegraphics[scale = 1.0]{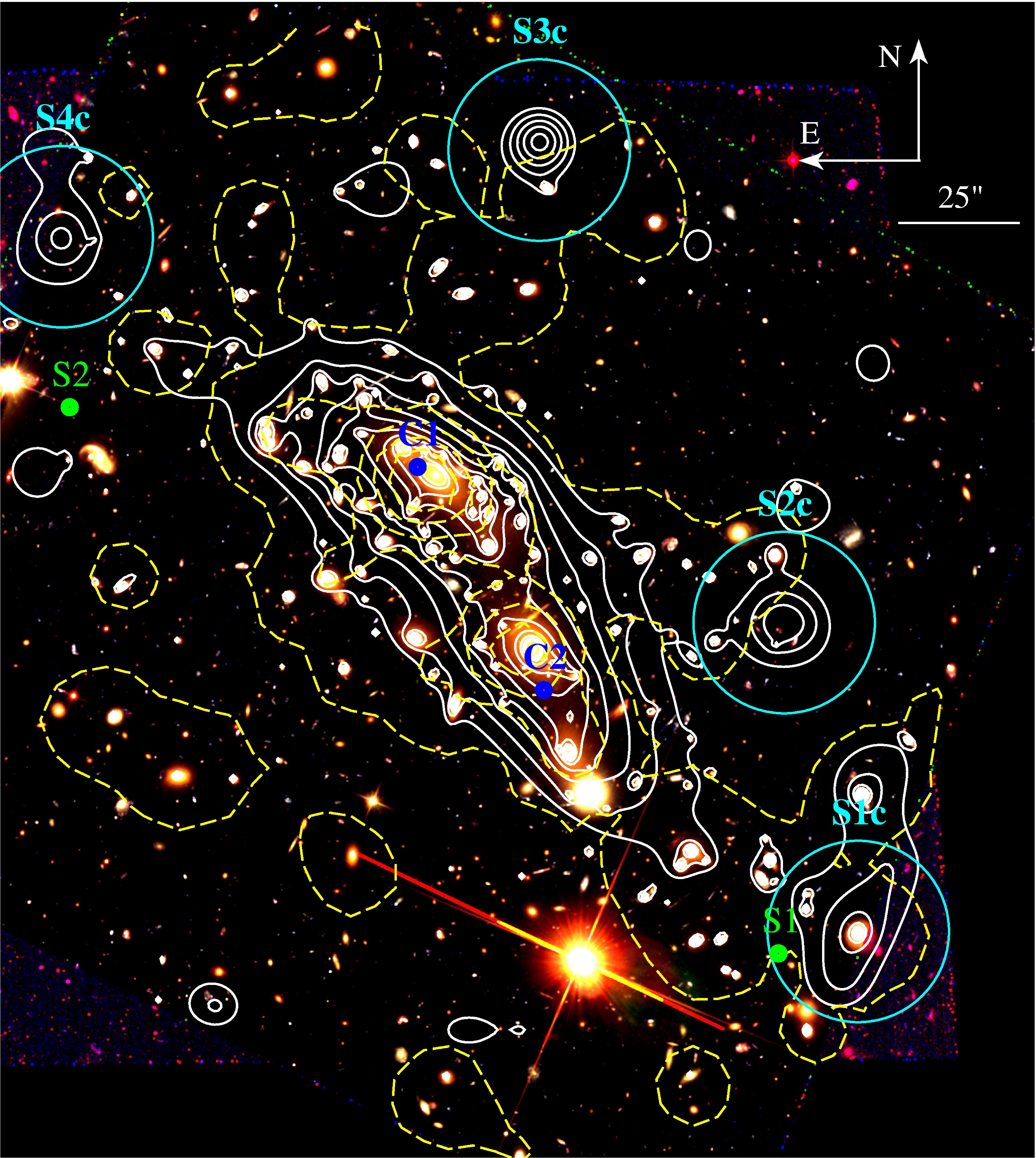}
    \caption{Composite colour \emph{HST}/ACS image using the $F814W$, $F606W$ and $F435W$ pass-bands, together with the projected mass (white solid line) and brightness contours (yellow dashed line) for the cluster field overlaid. Projected surface mass density contours are obtained from ($1.17 \times 10^{9}$) up to ($2.93 \times 10^{9}$)\,h$_{70}^{-1}$\,M$_{\odot}$\,kpc$^{-2}$. Brightness contours are obtained from $m_{F814W} = 23.3$ up to $m_{F814W} = 20.0$. We also mark the centres of the fixed halo components (C1 and C2), the detected substructures (S1 and S2) by \citetalias{Jauzac2015} and the detected substructure in this work, labelled as S1c, S2c, S3c and S4c. The size of the rings that enclose the substructures correspond to 100\,kpc at the cluster redshift. }
    \label{fig:main}
\end{figure*}

\begin{figure*}
    \centering
    \includegraphics[scale = 1.0]{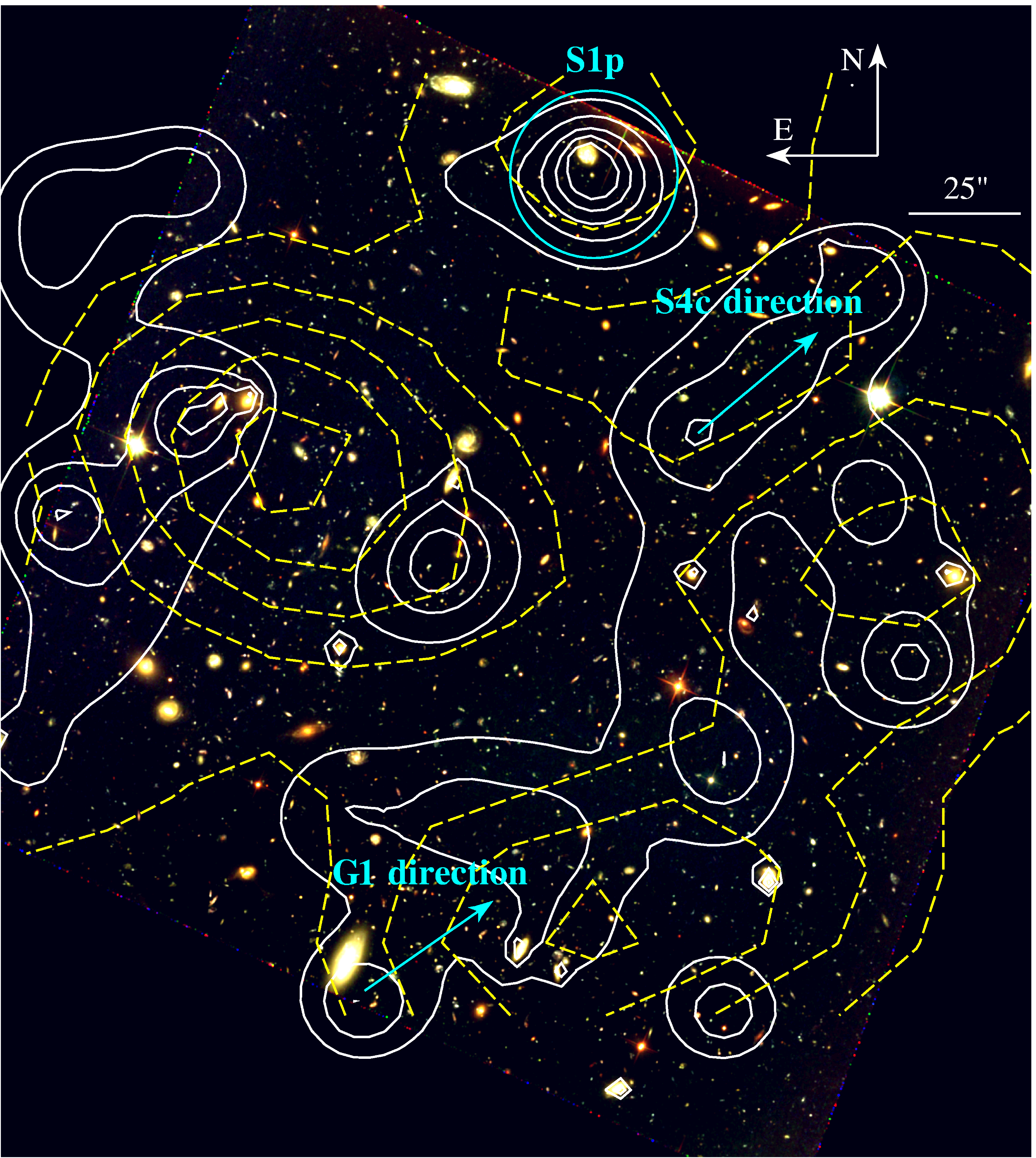}
    \caption{Composite colour \emph{HST}/ACS image using the $F814W$, $F606W$ and $F435W$ pass-bands, together with the projected mass (white solid line) and brightness contours (yellow dashed line) for the parallel field. Projected surface mass density contours are obtained from ($4.69 \times 10^{8}$) up to ($1.64 \times 10^{9}$)\,h$_{70}^{-1}$\,M$_{\odot}$\,kpc$^{-2}$.  Brightness contours are obtained from $m_{F814W} = 23.3$ up to $m_{F814W} = 20.8$. We mark the substructure detected in this work, S1p. The arrows indicate the directions to G1 and S4c.}
    \label{fig:parallel}
\end{figure*}

\subsection{Parallel field}

Figure~\ref{fig:parallel} shows the composite colour \emph{HST}/ACS image using the $F814W$, $F606W$ and $F435W$ pass-bands together with the projected mass and brightness contours for the parallel field. Projected surface mass density contours are obtained from ($4.69 \times 10^{8}$) to ($1.64 \times 10^{9}$)\,h$_{70}^{-1}$\,M$_{\odot}$\,kpc$^{-2}$. Brightness contours are obtained from  $m_{F814W} = 23.3$ to  $m_{F814W} = 20.8$. There is a general agreement between the projected mass and the brightness distributions particularly considering that the brightness map is poorly determined since it is based on only 26 cluster galaxies.

The overall projected surface mass density is consistent with a clumpy distribution connected by filament-like structures to the cluster core. In this field, we detect one substructure, S1p, located at RA\,(J2000) = 4:16:32.207, Dec\,(J2000) = $-$24:05:18.327, and with a projected mass within a $100$\,kpc aperture of $(2.1 \pm 0.4)\times 10^{13}$\,h$_{70}^{-1}$\,M$_{\odot}$. This is coincident with a peak in the brightness distribution, as well as with a galaxy overdensity. It appears to be a good correlation between the elongation of the projected mass distribution and the direction pointing to the cluster. In particular, two of the overdensities are aligned with the direction of S4c and G1 as marked in Fig. \ref{fig:main}. 

The detected substructures do not have evident counterparts in the \emph{Chandra} image of the field. We used \texttt{pyproffit}\footnote{\href{https://github.com/domeckert/pyproffit}{https://github.com/domeckert/pyproffit}} \citep{Eckert2011} to extract X-ray surface brightness profiles around the position of S1p. The X-ray signal is consistent with the background level. Assuming that the structure S1p is real and located at the redshift of the cluster, we can thus set an upper limit on the X-ray luminosity of $4.1\times10^{41}$ ergs/s ([0.5-2] keV rest frame, 90\% confidence level) within a circle of 1 arcmin radius around the source. However, we find a low-significance excess of X-ray emission located $\sim0.7$ arcmin from S1p, close to the arrow that points towards S4c in Fig. \ref{fig:parallel}. The tentative X-ray source is centred on RA\,(J2000) = 4:16:29.646, Dec=$-$24:05:44.044. We also extracted the brightness profile around this position, which confirms a 2.9$\sigma$ excess above the background. Again assuming that the source is located at the redshift of MACS\,J0416, we derive a luminosity of $(8.3 +/ 2.9)\times10^{41}$ erg/s in the [0.5-2] keV band (rest frame) within 1 arcmin radius. If the emission originates from a virialized infalling halo that has not yet interacted with MACS\,J0416, such a luminosity would be typical of a galaxy group with $kT\sim0.8$ keV and $M_{500c}\sim2\times10^{13}$ $M_\odot$ according to the scaling relations of \citet{Giles2016,Lieu2016}. Conversely, if S1p is confirmed to be a real substructure and given its substantially larger lensing mass, it would be almost entirely depleted of hot gas. This would imply that the surrounding dark matter halo has survived a previous interaction with the main cluster, whereas the gas content has been almost entirely stripped. According to the X-ray surface brightness profile of the main halo component C1, the core is composed of a very compact core and a more extended gas halo which suggest a possible previous merger event \citep{Ogrean2015} which would favour this scenario. 

\subsection{Substructure analysis}

According to the projected density distribution derived from the lensing analysis, we detect 5 substructures with $M$($R<200$\,kpc)$> 5\times10^{13}$\,h$_{70}^{-1}$\,M$_{\odot}$. The properties of massive substructures detected in massive galaxy clusters can be used as a test for $\Lambda$CDM, by comparing the observed detections with the subhalo mass function predicted by numerical simulations \citep[e.g. ][]{Natarajan2007, Grillo2015, Munari2016, Schwinn2017}. In order to do the comparison we select one cluster, with similar properties as MACS\,J0416, from the 30 Hydrangea/C-EAGLE simulated galaxy clusters described in Sect.\,\ref{sec:simu} and compare our lensing results with the subhalo distribution of the selected cluster within a projected 2D plane. 

\begin{figure}
    \centering
    \includegraphics[scale = 1.0]{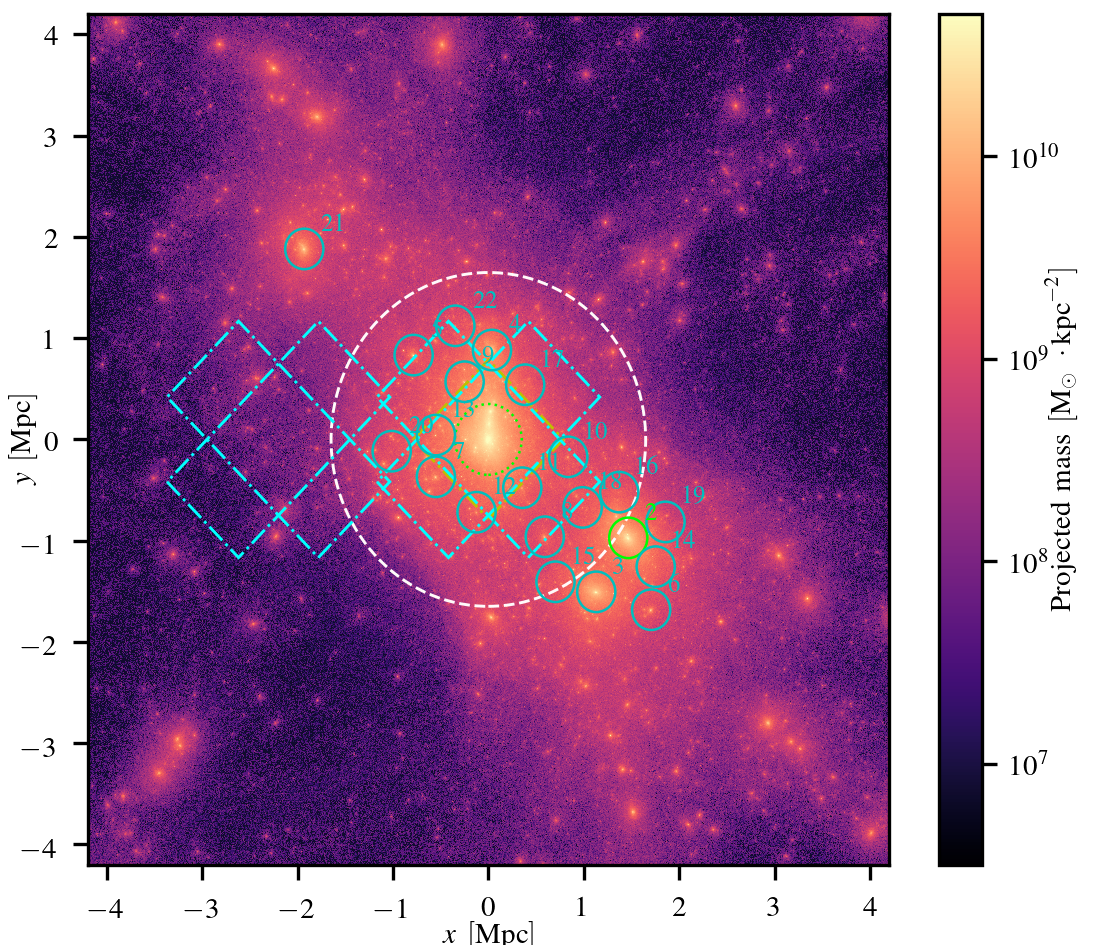}
    \caption{Projected mass map (using a logarithmic scale) of the selected simulated cluster at $z = 0.411$ with a $M_{200} = 7.43 \times 10^{14}\,{\rm M}_{\odot}$ and a  spherical over-density radius $R_{200} =1.64~{\rm Mpc}$  (dashed circle). The small cyan circles indicate the identified subhalos. Green circle labelled as 2, corresponds to the only subhalo with an aperture mass $> 5 \times 10^{13}\,M_{\odot}$, which is the expected detection threshold in our lensing analysis. The dotted circle of 350\,kpc radius corresponds to the central excluded region (see text for further details). Cyan dot-dash line is a representation for BUFFALO FOV.}
    \label{fig:simu}
\end{figure}

The selected simulated cluster is located at $z = 0.411$ with a $M_{200} = 7.2 \times 10^{14}\,M_{\odot}$ \citep[Cluster ID: CE-28, according to Table\,A1 in][]{Barnes2017}. This corresponds to projected aperture masses: $M$($R<200$\,kpc) $= 1.5 \times 10^{14}\,M_{\odot}$, $M$($R<250$\,kpc) $= 1.9 \times 10^{14}\,M_{\odot}$ and $M$($R < 320$\,kpc) $ = 2.50 \times  10^{14}\,M_{\odot}$. In order to identify subhalos\footnote{Here we refer to all identified halos within the considered region as subhalos, even if they are not included within the main halo of the simulated cluster.} in the field that would be detected in our lensing analysis, we first select objects from the subhalo catalogue within a 3\,Mpc radius, excluding a central region of 350\,kpc to mimic the lack of sensitivity to dense structures due to the presence of the dense core. Then we project their centres into the sky plane within $\pm 5$\,Mpc, and compute the projected masses enclosed by circular apertures of $200$\,kpc radius. Detected subhalos are marked in Fig. \ref{fig:simu} together with the projected mass map of the simulated cluster. We only detect one subhalo with an aperture mass $> 5 \times 10^{13}\,M_{\odot}$, which is the expected detection threshold in our lensing analysis. This subhalo, labelled as 2 in Fig. \ref{fig:simu}, has an aperture mass of $5.7 \times 10^{13}\,M_{\odot}$ and is located at 1703\,kpc from the cluster centre, at a similar distance as the substructure reported in the parallel field, S1p. However, the selected parallel field is located perpendicular to the mass distribution while the subhalo is located in a direction close to the cluster elongation. 

It is important to consider that the HFF fields only represent $\sim$20\% of the area considered for the identification of subhalos in the simulated cluster. Moreover, the observed region is only continuous out to $\sim 800$\,kpc from the cluster centre, where four of the five reported substructures are located. Within this region, we identify six subhalos in the simulated cluster, with an average mass $\langle M(R < 200$\,kpc$) \rangle = 2 \times 10^{13}\,M_{\odot}$, the most massive one located at 600\,kpc from the centre with $M(R < 200$\,kpc$) = 3.1 \times 10^{13}\,M_{\odot}$. Therefore, we can argue that there are significant differences between the subhalo distribution observed in the simulated cluster and the one observed in MACSJ\,0416. Since the simulated cluster appears to be relatively dynamically relaxed, and was observed as an isolated halo, it is sensible to think we are in the case of a more evolved system than MACS\,J0416 itself. MACS\,J0416 is a very elongated cluster, showing an obvious bimodal density mass distribution. Taking into account that subhalos tend to fall in the inner cluster regions at lower redshifts, the simulated cluster could be representing the next evolutionary stage of MACS\,J0416 in which the closest subhalos have already merged with the cluster core. Discrepancies between the observed radial distribution of subhalos and the one simulated were already reported by \citet{Natarajan2017}. One reason could be that the selected simulated cluster is not representative of MACS\,J0416. HFF clusters were selected for their strong magnifying power, which biases the selection towards dynamically complex and extremely massive systems.

\begin{figure*}
    \centering
    \includegraphics[scale = 0.6]{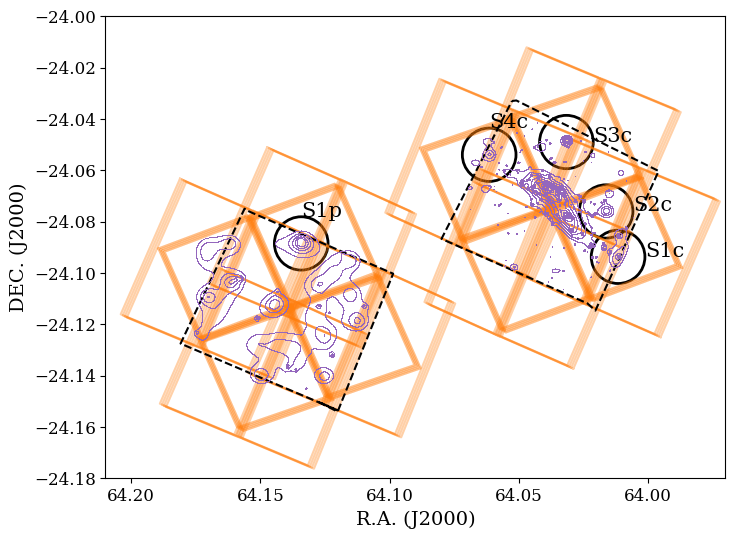}
    \caption{Substructure locations within the HFF (dashed black lines) and BUFFALO (orange lines) footprints. The circles correspond to an aperture of 200\,kpc radius. Contours correspond to the derived projected density distributions in the parallel and cluster field. It is worth noting that these contours are obtained at different density levels as specified in Fig. \ref{fig:main} and \ref{fig:parallel}.}
    \label{fig:footprint}
\end{figure*}

It is worth noting that three of the reported MACSJ\,0416 substructures in this work are located close to the edges of the HFF field of view. This can significantly hamper the mass estimates of the detected substructures. The BUFFALO survey will triple the observed area, providing an almost continuous region between the cluster and the parallel fields (Fig. \ref{fig:footprint}). It would thus provide a major improvement in the aperture masses estimates. Although the observed depth will be lower than the HFF observations (the exposure time for HFF is around 140 HST orbits while for BUFFALO is going to be of 4, where each orbit corresponds to 2028\,s), BUFFALO is expected to detect the substructures reported in the cluster field and, therefore, to better characterise their physical properties. 

\section{Summary and conclusion}
\label{sec:conclusions}

In this work we present the analysis of the matter distribution in the HFF massive galaxy cluster, MACS\,J0416. The analysis includes an optical analysis of the cluster core (cluster field), as well as its adjacent HFF `blank' field (parallel field) combined with an X-ray study. This work is motivated by the upcoming observations from the BUFFALO survey, which will complete the region between the analysed fields.

We derive the projected surface mass density obtained from our lensing analysis, and compare our results with the optical and X-ray emission distributions. For both fields there is a good agreement between the projected density distribution and the optical emission. The resulting total masses computed within circular apertures centred on the brightest galaxy member of the cluster are higher than previous determinations \citep{Gruen2014,Johnson2014,Richard2014,Jauzac2014,Jauzac2015,Grillo2015}. 
Discrepancies in these mass estimates and derived projected mass density distributions could be due to the different datasets, the different assumptions used by different lensing mass modeling algorithms, and the fact that the mass-sheet degeneracy is only partially broken by the inclusion of photometric redshifts \citep{Grillo2015}. \citet{Meneghetti2017} compared different lens modeling techniques to derive magnification estimates for the same clusters, and obtained very good agreements between the different techniques. Moreover, derived magnification maps by different authors led to similar parameters to calibrate the luminosity function \citep{Ishigaki2018}. This shows that the obtained overall projected mass density distribution is in good agreement with previous analyses. Also, \citet{Remolina2018} evaluate the predictive power of strong lensing models for MACS\,J0416, obtaining a good agreement in the arc prediction of the considered models. Although lensing modelling techniques have proven to be accurate for reconstructing projected density distributions, analyses based on different datasets, which include different redshift information, can lead to discrepancies in mass estimates. Therefore, more accurate error estimates that take into account these potential biases can be important in order to derive total masses values.   

We identify five substructures in both fields at the $>5\sigma$ level. Four of them are located in the cluster field with one of them matching a detection previously reported by \citetalias{Jauzac2015}. 
The identified substructures are also detected in the surface density profiles, when they are computed in triangular regions pointing to each of them. Three of the detected substructures, S1c, S2c and S1p, lie close to galaxy overdensities which reinforces their identification. Moreover, S2c is located close to an excess in the X-ray emission according to the map presented by \citet{Ogrean2015}. In the case of S3c and S4c, we suspect that these structures can be generated by the imposition of the parametric model in the mass modelling, since they are not detected if we only use the grid modelling for the surface projected mass reconstruction. 

For the parallel field, we obtain a clumpy projected mass distribution connected by filament-like structures. The overall projected mass distribution shows a potential alignement with the cluster direction, since two of the overdensities are elongated pointing to the brightest member galaxy and to one of the detected substructures in the cluster field. This is a key result because this field was selected expecting no significant mass distribution associated with the cluster. The detected substructure in this field, S1p, has no evident X-ray counterpart. The lensing mass estimated for this structure might suggest that it previously interacted with MACS\,J0416 and, as a result of this interaction, almost the whole gas content was stripped away. In this scenario, the detected low-significance excess of X-ray emission at $\sim0.7$ arcmin from S1p could be associated with the stripped gas as previously observed in other galaxy systems at low redshift \citep[eg.][]{Eckert2017}. Nevertheless, this scenario should be less common whithout a remnant core. It is important to take into account that S1p is close to the edge of the FOV, thus border issues can considerably hamper the determination of the substructure physical properties such as the aperture mass and exact location. Further studies based on BUFFALO observations can confirm the presence of this structure and reinforce the striping scenario. 

In order to test our results and make further predictions for the BUFFALO survey, we compare the distribution of substructures in MACS\,J0416 with the one observed in a Hydrangea/C-EAGLE simulated cluster. We identify 21 subhalos within a 3.0\,Mpc radius from the cluster centre. Only one of these subhalos has an aperture mass $M$($R<200$\,kpc) $>5\times10^{13}$\,M$_{\odot}$, which is the expected threshold according to our lensing analysis. This subhalo is located at a distance centre that is in agreement with that observed for S1p, but in a direction close to the simulated cluster elongation. In the inner region ($< 800$\,kpc) of the simulted cluster, where we report 4 substructers for MACS\,J0416, none of the identified subhalos satisfy the lensing aperture mass threshold, with the most massive identified subhalo within this region having $M(R < 200$\,kpc$) = 3.1 \times 10^{13}\,M_{\odot}$. Therefore, we conclude that the simulated cluster represents a dynamically more evolved system in which all of the subhalos close to the core have already merged with the cluster.  Discrepancies in the radial distribution of subhalos may be due to the fact that the simulated cluster does not adequately reproduce the observational properties of the HFF clusters, since the selection criteria of the HFF systems can introduce bias towards massive and complex cluster systems. In fact, MACS\,J0416 shows a very elongated and bimodal mass distribution which is not the case for the selected simulated cluster.

BUFFALO will be of a major importance to confirm and characterise the substructures detected in the cluster field, mainly those close to the edges of the FOV, and thus understand better the build-up and merging scenario in place in MACS\,J0416. Moreover, the overall projected density distribution of the parallel field seems to be connected with the cluster. BUFFALO data will link both fields, and will thus shed light on this possible connection. This work is just a glimpse into the promising data that future surveys will provide in order to strengthen our understanding of these giant galaxy cluster systems.

\section*{Acknowledgements}
We kindly thanks Liliya Williams and Johannes Schwinn for their comments that improved the work. 
This work is based on data and catalogue products from HFF-DeepSpace, funded by the National Science Foundation and Space Telescope Science Institute (operated by the Association of Universities for Research in Astronomy, Inc., under NASA contract NAS5-26555).
This project has received funding from the European Union's Horizon 2020 Research and Innovation Programme under the Marie Sklodowska-Curie grant agreement No 734374.
This work was partially supported by the Consejo Nacional de Investigaciones Cient\'{\i}ficas y T\'ecnicas (CONICET, Argentina) 
and the Secretar\'{\i}a de Ciencia y Tecnolog\'{\i}a de la Universidad Nacional de C\'ordoba (SeCyT-UNC, Argentina).
We made an extensive use of the following python libraries:  http://www.numpy.org/, http://www.scipy.org/, and http://www.matplotlib.org/. This research made use of Astropy,\footnote{http://www.astropy.org} a community-developed core Python package for Astronomy \citep{astropy:2013, astropy:2018}. 
MJ is supported by the United Kingdom Research and Innovation (UKRI) Future Leaders Fellowship 'Using Cosmic Beasts to uncover the Nature of Dark Matter' (grant number MR/S017216/1). This project was also supported by the Science and Technology Facilities Council [grant number ST/L00075X/1]. DH is supported by the D-ITP consortium, a program of the Netherlands Organization for Scientific Research (NWO) that is funded by the Dutch Ministry of Education, Culture and Science (OCW). MS is supported by the Netherlands Organization for Scientific Research (NWO) VENI grant 639.041.749.


\bibliography{references}

%%%%%%%%%%%%%%%%%%%%%%%%%%%%%%%%%%%%%%%%%%%%%%%%%%

% Don't change these lines
\bsp	% typesetting comment
\label{lastpage}
\end{document}